\documentstyle[12pt]{article}

\global\arraycolsep=2pt

\begin{document}

\begin{titlepage}

\begin{flushright}
CERN-TH.7087/93\\
November 1993
\end{flushright}

\vspace{0.3cm}

\begin{center}
\Large\bf QCD-Based Interpretation of the\\
Lepton Spectrum\\
in Inclusive $\bar B\to X_u\,\ell\,\bar\nu$ Decays
\end{center}

\vspace{0.8cm}

\begin{center}
Matthias Neubert\\
{\sl Theory Division, CERN, CH-1211 Geneva 23, Switzerland}
\end{center}

\vspace{0.8cm}

\begin{abstract}
We present a QCD-based approach to the endpoint region of the lepton
spectrum in $\bar B\to X_u\,\ell\,\bar\nu$ decays. We introduce a
genuinely nonperturbative form factor, the shape function, which
describes the fall-off of the spectrum close to the endpoint. The
moments of this function are related to forward scattering matrix
elements of local, higher-dimension operators. We find that
nonperturbative effects are dominant over a finite region in the
lepton energy spectrum, the width of which is related to the kinetic
energy of the $b$-quark inside the $B$ meson. Applications of our
method to the extraction of fundamental standard model parameters,
among them $V_{ub}$, are discussed in detail.
\end{abstract}

\centerline{(submitted to Physical Review D)}

\end{titlepage}

\section{Introduction}

Recently, much progress has been achieved in the understanding of
inclusive weak decays of hadrons containing a heavy quark $Q$. Using
the theoretical tools of the operator product expansion and the heavy
quark effective theory (HQET) \cite{Eich,Geor,Mann,Falk,habil}, one
can construct a systematic expansion of the (differential) decay
distributions in powers of $\Lambda/m_Q$, where $\Lambda$ is a
characteristic low-energy scale of the strong interactions
\cite{Chay,Bigi,Blok,MaWe,Thom}. Quite remarkably, the parton model
emerges as the leading term in this QCD-based expansion, and the
nonperturbative corrections to it are suppressed by a factor
$\Lambda^2/m_Q^2$. The fact that there are no first-order power
corrections relies on a particular definition of $m_Q$, which is
provided in a natural way by requiring that there be no residual mass
term for the heavy in HQET \cite{AMM,SR3}. This definition is unique
and can be regarded as a nonperturbative generalization of the
concept of a pole mass.

The availability of a systematic, QCD-based expansion of inclusive
decay rates raises the hope for a better understanding of these
processes in general, and in particular for a more reliable
extraction of the standard model parameters $m_b$, $m_c$, $V_{cb}$,
and $V_{ub}$, which was so far hindered by strong model dependence.
For a determination of $V_{ub}$, however, it is essential to
understand the endpoint region of the lepton spectrum, which is of
genuinely nonperturbative nature. Although the new methods developed
in Refs.~\cite{Bigi,Blok,MaWe,Thom} provide an important step towards
this goal, they are not directly applicable to the endpoint region.
The difficulties arise from the fact that close to the endpoint the
expansion parameter is no longer $\Lambda/m_b$, but $\Lambda/
(m_b-2 E_\ell)$, and thus the theoretical prediction becomes singular
when the lepton energy approaches the parton model endpoint
$E_{\ell,{\rm max}}=m_b/2$. It is then not obvious how to interpret
the theoretical predictions.

To see what the problem is, consider the theoretical prediction for
the lepton spectrum in $\bar B\to X_u\,\ell\,\bar\nu$ decays.
Including the leading nonperturbative corrections, one obtains
\cite{Bigi,Blok,MaWe}
\begin{equation}\label{spectrum}
   {1\over 2\Gamma_b}\,{{\rm d}\Gamma\over{\rm d}y}
   = y\,F(y)\,\Theta(1-y) - {\lambda_1+33\lambda_2\over 6 m_b^2}\,
   \delta(1-y) - {\lambda_1\over 6 m_b^2}\,\delta'(1-y) \,,
\end{equation}
where
\begin{equation}
   y = {2 E_\ell\over m_b} \,,\qquad
   \Gamma_b = {G_F^2\,|\,V_{ub}|^2\over 192\pi^3}\,m_b^5 \,,
\end{equation}
and

\begin{equation}
   F(y) = (3-2y)\,y + {5y^2\over 3}\,{\lambda_1\over m_b^2}
   + (6+5y)\,y\,{\lambda_2\over m_b^2} \,.
\end{equation}
For simplicity, we do not include perturbative QCD corrections, which
have been calculated in Refs.~\cite{Cabi,ACM}. The hadronic
parameters $\lambda_1$ and $\lambda_2$ are related to the kinetic
energy $K_b$ of the heavy quark inside the $B$ meson, and to the mass
splitting between $B$ and $B^*$ mesons \cite{FaNe}:
\begin{equation}\label{lamdef}
   K_b = -{\lambda_1\over 2 m_b} \,,\qquad
   m_{B^*}^2 - m_B^2 = 4\lambda_2 \,.
\end{equation}
The singular structure of the operator product expansion close to the
endpoint at $y=1$ manifests itself in the appearance of
$\delta$-function (and higher) distributions. Certainly, one cannot
trust the shape of the theoretical spectrum close to the endpoint.
Nevertheless, integrating (\ref{spectrum}) with a smooth weight
function, one obtains well-behaved results for quantities such as the
total decay rate and the average lepton energy:
\begin{equation}
   \Gamma = \Gamma_b\,\bigg\{ 1
   + {\lambda_1-9\lambda_2\over 2 m_b^2} \bigg\} \,,\qquad
   \langle E_\ell\rangle = {7 m_b\over 20}\,\bigg\{ 1
   - {7\lambda_1+57\lambda_2\over 14 m_b^2} \bigg\} \,.
\end{equation}
The coefficients of the singular terms give nonvanishing
contributions to these integrated quantities. There is thus some
relevant physical information contained in these terms.

Bigi et al.\ \cite{Bigi} have advocated to integrate over the
singularities before confronting the theoretical prediction for the
lepton spectrum with data. They argued that one has to integrate over
a finite energy interval of at least several hundred MeV,
corresponding to a region of order $1/m_b$ in the variable $y$.
This proposal is based on the idea of quark-hadron duality, which
implies that when one sums over a sufficient number of exclusive
hadronic modes, the decay probability into hadrons equals the decay
probability into free quarks.\footnote{One expects that duality holds
for the lepton spectrum even in the endpoint region, which extends
over an interval of order $1/m_b$ in $y$. Only in a tiny region of
order $1/m_b^2$ below the physical endpoint, the spectrum is
dominated by a few exclusive modes.}
Note that the $\delta$-function term in (\ref{spectrum}) contributes
to the integrated spectrum, but the term proportional to
$\delta'(1-y)$ does not. Similarly, more singular terms, which appear
at higher orders in the $1/m_b$ expansion, do not contribute.

A slightly modified procedure was proposed by Manohar and Wise
\cite{MaWe}. They chose to smear the spectrum with a gaussian
distribution of width $\Delta y$. Empirically, they found that
$\Delta y\sim 0.2-0.5$ is necessary to obtain from the theoretical
prediction a smooth lepton spectrum, which can be compared to data.
This procedure has the disadvantage that the results depend on the
smearing function, and that the choice of $\Delta y$ is ad hoc. When
the smearing function is chosen to be symmetric, it again follows
that the term proportional to $\delta'(1-y)$ does not contribute to
the smeared spectrum.

The frustrating conclusion of these analyses is that the new
theoretical methods are only of very limited use for a more reliable
determination of $V_{ub}$, since the region of the lepton spectrum
which is accessible to a measurement is smaller than the region over
which the theoretical spectrum has to be integrated in order to
obtain a reasonable result.

As proposed in Refs.~\cite{Bigi,Blok,MaWe}, the theoretical
description is to a large extent ignorant to the rich physical
information contained in the lepton spectrum close to the endpoint.
In this paper, we shall suggest a different approach. It is motivated
by a very simple observation: In the parton model, the endpoint
region of the lepton spectrum is described by a step function, the
location of which is determined by the kinematics of a free quark
decay. The true physical endpoint, however, is determined by the
decay kinematics of hadrons. Hence, when QCD is trying to tell us
something about the redistribution of the endpoint region due to
nonperturbative effects, it can only do this by the occurance of
singular functions. Our approach will allow us to extract the
physical information contained in the singular terms in the
QCD-predicted lepton spectrum in a systematic way, and to all orders
in the $1/m_b$ expansion. To this end, we shall introduce the concept
of a {\sl shape function\/}, which is a genuinely nonperturbative
form factor that describes the fall-off of the spectrum in the
endpoint region. We find that the characteristic width of this region
is given by $\sigma_y=(-\lambda_1/3 m_b^2)^{1/2}$, corresponding to a
{\sl finite\/} region in the lepton energy. Although there do not
appear first-order power corrections in (\ref{spectrum}), it is
important to realize that there exists a small region where the true
spectrum is very different from the theoretical prediction. This
difference is described by the shape function. We will show that the
{\sl moments\/} of this function can be addressed in QCD, and can be
related to hadronic parameters (such as $\lambda_1$ and $\lambda_2$)
that are defined in terms of forward scattering matrix elements of
local, higher-dimension operators. To all orders in $1/m_b$, the
leading contributions to the moments can be given in closed form.

We believe that our approach will eventually lead to a better
understanding of the nonperturbative aspects of the lepton spectrum
close to the endpoint. It establishes the connection between the
experimentally observed lepton spectrum and the underlying theory of
QCD. This connection works in both directions: Theoretical ideas
about the moments of the shape function can help to analyze the
lepton spectrum and to determine the values of the quark masses and
mixing angles. On the other hand, from a precise measurement of the
spectrum in the endpoint region one can extract the shape function
and with it some fundamental matrix elements of higher-dimension
operators in QCD.

Starting from a resummation of the theoretical lepton spectrum, we
pre\-sent in Sec.~\ref{sec:2} a heuristic argument that leads to the
notion of a function $S(y)$, which describes the fall-off of the
spectrum in the endpoint region. In Sec.~\ref{sec:3}, we introduce
this shape function, discuss its properties, and derive relations for
the first two moments of $S(y)$. Sec.~\ref{sec:4} is devoted to a
formal definition of the shape function to all orders in $1/m_b$. The
leading contributions to the moments are related to forward
scattering matrix elements of local, higher-dimension operators in
HQET. For the purpose of illustration, a simple model calculation of
the shape function is presented in Sec.~\ref{sec:5}. In
Sec.~\ref{sec:6}, we summarize our results, indicate possible further
applications and improvements of the method, and give some
conclusions.

\section{Resummation of the singular terms}
\label{sec:2}

To motivate the concept of a shape function, let us try to resum the
singular contributions in (\ref{spectrum}) into a corrected parton
model decay distribution. Obviously, the term proportional to
$\delta(1-y)$ can be absorbed by a shift of the argument of the step
function in the leading-order term. More interesting is the
contribution proportional to $\delta'(1-y)$. It arises at second
order in the expansion of the step function. However, since there is
no $\delta$-function contribution of first order in $1/m_b$, it
follows that one needs more than one step function, resulting in a
{\sl dispersion\/} of the spectrum. In fact, to order $1/m_b^2$, we
can rewrite the theoretical spectrum in the following way:
\begin{equation}\label{resum}
   {1\over 2\Gamma_b}\,{{\rm d}\Gamma\over{\rm d}y} = y\,F(y)\,
   {1\over N}\,\sum_{i=1}^N\,\Theta(1-y+\varepsilon_i) \,,
\end{equation}
where
\begin{equation}\label{constr}
   \delta y = {1\over N}\,\sum_{i=1}^N\,\varepsilon_i
   = -{\lambda_1+33\lambda_2\over 6 m_b^2} \,,\qquad
   \sigma_y^2 = {1\over N}\,\sum_{i=1}^N\,\varepsilon_i^2
   = -{\lambda_1\over 3 m_b^2} \,.
\end{equation}
{}From the second relation it follows that the displacements
$\varepsilon_i$ are of order $1/m_b$. Thus, the first relation
corresponds to a nontrivial cancellation. Note that $\sigma_y$ can be
identified with the characteristic width of the endpoint region,
i.e., the region over which the spectrum is dominated by
nonperturbative effects. This is already a remarkable conclusion: The
coefficient of the most singular term in (\ref{spectrum}), which had
no effect in the approaches of Refs.~\cite{Bigi,MaWe}, determines the
size of the endpoint region.

At this point, it is worthwhile to obtain some estimates of the
nonperturbative corrections. From the observed value of the mass
splitting between $B$ and $B^*$ mesons one obtains $\lambda_2\simeq
0.12$ GeV$^2$. The parameter $\lambda_1$, on the other hand, is not
directly related to an observable. Recently, we have shown that the
field-theory analog of the virial theorem relates the kinetic energy
of a heavy quark inside a hadron (and thus $\lambda_1$) to a matrix
element of the gluon field strength tensor \cite{virial}. This
theorem makes explicit an intrinsic ``smallness'' of $\lambda_1$,
which was not taken into account in existing QCD sum rule
calculations of this parameter \cite{Subl,Elet,BaBr}. As a
consequence, we expect that $(-\lambda_1)$ is considerably smaller
than predicted in these analyses. Here we shall use the range
$-\lambda_1=0.05-0.30$ GeV$^2$. According to its definition,
$\lambda_1$ is negative, so that the width $\sigma_y$ in
(\ref{constr}) is well defined. Using these numbers, as well as
$m_b=4.8$ GeV, we estimate $\delta y \simeq-0.03$ and $\sigma_y\simeq
0.03-0.07$. We can multiply these quantities by $m_b/2$ to obtain the
corresponding shift and spread in the lepton energy spectrum. They
are $\delta E\simeq -65$ MeV and $\sigma_E=(-\lambda_1/12)^{1/2}
\simeq 65-160$ MeV. The value of $\sigma_E$ can be compared to the
width of the gap between the parton model endpoint of the spectrum
and the physical endpoint, which, if we neglect the pion mass, is
located at $E_{\ell,\rm max}^{\rm phys}=m_B/2$. This width is $\Delta
E\simeq (m_B-m_b)/2\simeq 240$ MeV. These numbers seem quite
reasonable. In fact, assuming that the distribution of the
displacements $\varepsilon_i$ around $y=1$ is approximately
symmetric, we have to require that $\sigma_E<\Delta E$, which is
equivalent to $-\lambda_1<3 (m_B-m_b)^2$. For reasonable values of
$\lambda_1$, this bound is always satisfied.

In Fig.~\ref{fig:1}, we show the resummed lepton energy spectrum
(\ref{resum}) for $\lambda_1=-0.2$ GeV$^2$, $\lambda_2=0.12$ GeV$^2$,
$N=10$, and a particular set of $\varepsilon_i$ satisfying the
constraints in (\ref{constr}). For this choice of parameters, the
dispersion in the spectrum is such that the endpoint falls close to
the physical endpoint at $y\simeq 1.1$. Our reinterpretation of the
QCD-predicted lepton spectrum has led to a reasonable shape which,
however, is quite arbitrary. In fact, increasing $N$, we can generate
any decreasing function that satisfies the constraints in
(\ref{constr}). In the following section, we will introduce a shape
function $S(y)$ instead of the sum over step functions. The
constraints will then turn into predictions for the first two moments
of this function.

\section{The shape function}
\label{sec:3}

We proceed by replacing the sum over step functions in (\ref{resum})
by a continuous function $\vartheta(y)$, which we furthermore
decompose as $\vartheta(y) = \Theta(1-y) + S(y)\,F(1)/F(y)$. The form
of the second term is chosen for later convenience. We shall refer to
$S(y)$ as the shape function. The support of this function is
restricted to a small interval of width $2\Delta$ around $y=1$, where
$\Delta$ of order $1/m_b$.\footnote{More precisely, we should require
that $S(y)$ be exponentially small outside an interval of width
$2\Delta$.}
Some basic properties of $S(y)$ can be derived from the physical
requirements that the differential decay rate by positive, and that
$\vartheta(y)$ be a continuous function. We note that
\begin{eqnarray}
   S(y) &=& 0\quad {\rm if}\quad |y-1|>\Delta \,, \nonumber\\
   S(y) &\ge& 0\quad {\rm if}\quad y\ge 1 \,, \nonumber\\
   \lim_{\epsilon\to 0}\,S(1+\epsilon) &-& S(1-\epsilon) = 1
    \,, \nonumber\\
   \lim_{\epsilon\to 0}\,S'(1+\epsilon) &-& S'(1-\epsilon) = 0 \,.
\end{eqnarray}
Moreover, we expect that $S'(y)\le 0$ if $y\ne 1$, but we shall not
impose this as a condition on $S(y)$.

As emphasized in Refs.~\cite{Bigi,MaWe}, because of the singular form
of the operator product expansion one has to integrate the
theoretical lepton spectrum with a smooth function before it can be
compared to data. On the set of smooth functions (i.e., functions of
$y$ which are slowly varying over scales of order $1/m_b$), a rapidly
varying function such as $S(y)$, which vanishes outside a small
interval around $y=1$, obeys a singular expansion of the
form\footnote{This procedure is familiar from the multipole expansion
of a localized distribution of charges in electrodynamics.}
\begin{equation}\label{singular}
   S(y) = \sum_{n=0}^\infty\,{M_n\over n!}\,\delta^{(n)}(1-y) \,,
\end{equation}
where the {\sl moments\/} $M_n$ are defined as
\begin{equation}\label{Mndef}
   M_n = \int\limits_0^\infty\!{\rm d}y\,(y-1)^n\,S(y)
   = \int\limits_{-\Delta}^\Delta\!{\rm d}y\,(y-1)^n\,S(y) \,.
\end{equation}
To see that (\ref{singular}) is correct, assume that any reasonable
test function $f(y)$ can be Taylor-expanded around $y=1$, with the
result that
\begin{equation}
   \int\limits_0^\infty\!{\rm d}y\,f(y)\,S(y)
   = \sum_{n=0}^\infty\,{M_n\over n!}\,f^{(n)}(1) \,.
\end{equation}

In terms of the shape function, the theoretical lepton spectrum takes
the form
\begin{equation}\label{result}
   {1\over 2\Gamma_b}\,{{\rm d}\Gamma\over{\rm d}y}
   = y\,\Big[ F(y)\,\Theta(y) + F(1)\,S(y) \Big] \,.
\end{equation}
{}From a comparison with (\ref{spectrum}), we find that the first two
moments of $S(y)$ must satisfy\footnote{Our judicious choice of the
prefactor $y$ in front of $S(y)$ has eliminated the contribution of
$\lambda_1$ to $M_0$.}
\begin{eqnarray}\label{M0M1val}
   M_0 &=& -{11\lambda_2\over 2 m_b^2} \simeq -2.9\% \,, \nonumber\\
   M_1 &=& -{\lambda_1\over 6 m_b^2} = {\sigma_y^2\over 2}
    \simeq (0.4-2.2)\times 10^{-3} \,.
\end{eqnarray}
Notice that $M_n={\cal{O}}(1/m_b^{n+1})$ by dimensional analysis.
Hence, the QCD prediction that $M_0={\cal{O}}(1/m_b^2)$ corresponds
to a nontrivial cancellation: The area under the shape function
(almost) vanishes. The first moment, which is related to the
characteristic width of the endpoint region, is of the expected order
of magnitude.

The concept of a shape function exploits to full extent the physical
information contained in the coefficients of the singular terms in
the QCD-predicted lepton spectrum. We find that over a region of
width $2\Delta\propto 1/m_b$, the spectrum is of genuinely
nonperturbative nature and described by a function $S(y)$, the
moments of which can be addressed in QCD. When one goes to higher
orders in the $1/m_b$ expansion, one can address higher moments. In
fact, the moments obey an expansion of the form
\begin{equation}\label{Mnexp}
   M_n = {a_n\over m_b^{n+1}} + {b_n\over m_b^{n+2}} + \ldots,
\end{equation}
where so far we know the coefficients $a_0=0$, $b_0=-11\lambda_2/2$,
and $a_1=-\lambda_1/6$. With the exception of the moment $M_0$, where
the leading term $a_0$ vanishes, we may argue that it would be a good
approximation to know the leading coefficient $a_n$ for each moment.
The corrections involving $b_n$ only change the moments by small
amounts. On the other hand, knowledge of a new moment teaches us a
new piece of essential information about the shape of the spectrum in
the endpoint region. The higher moments give a small contribution to
integrated quantities such as the total decay rate, simply because in
(\ref{Mndef}) one integrates over a small region. Nevertheless, they
can affect the shape of the endpoint region in a substantial way.
What is relevant to the shape are the rescaled moments
\begin{equation}\label{calMn}
   {\cal{M}}_n = \int\limits_0^\infty\!{\rm d}E_\ell\,
   (2 E_\ell-m_b)^n\,S(E_\ell) = m_b^{n+1}\,M_n
   = a_n + {b_n\over m_b} + \ldots,
\end{equation}
which remain nonzero in the limit $m_b\to\infty$. As an illustration
of the importance of higher moments, we show in Fig.~\ref{fig:2} two
shape functions which have the same first two moments $M_0$ and
$M_1$, but different third (and higher) moments. The total decay rate
and the average lepton energy are the same in both cases (up to terms
of order $1/m_b^3$), but obviously the behavior close to the endpoint
is quite different.

\section{Formal definition of the shape function}
\label{sec:4}

The above discussion shows that for an understanding of the lepton
spectrum in the endpoint region, it is insufficient to truncate the
theoretical calculation at order $1/m_b^2$. Instead, what one needs
to investigate to all orders in $1/m_b$ are the coefficients $a_n$ in
(\ref{Mnexp}) and (\ref{calMn}). They arise from the most singular
terms in the shape function. In this section, we give a formal
definition of these terms to all orders in $1/m_b$. This will provide
us with a relation between the coefficients $a_n$ and forward
scattering matrix elements of local, higher-dimension operators in
HQET.

As mentioned in the introduction, the derivation of the lepton
spectrum is based on the operator product expansion in connection
with an expansion of hadronic matrix element in powers of $1/m_b$, as
provided by HQET. This is explained in detail in
Refs.~\cite{Bigi,Blok,MaWe,Thom}. Using the same technology, we can
derive a closed expression for the most singular terms of the shape
function, where $S(y)$ is defined as in (\ref{result}) as the sum of
all terms in the theoretical spectrum that become singular in the
limit $y\to 1$. We obtain the formal result
\begin{equation}\label{Sdef}
   S(y) = \bigg\langle \Theta\bigg[ 1 - y + {2\over m_b}\,
   (v-\hat p)\cdot iD\bigg] - \Theta(1-y) \bigg\rangle
   +~\hbox{less singular terms,}
\end{equation}
which is valid to all orders in the $1/m_b$ expansion. Here

$\hat p=p_\ell/m_b$ denotes the rescaled lepton momentum, and we
define the expectation value of an operator $O$ as
\begin{equation}
   \langle\,O\,\rangle =
   {\langle B(v)|\,\bar h_v\,O\,h_v\,|B(v)\rangle \over
    \langle B(v)|\,\bar h_v\,h_v\,|B(v)\rangle} \,.
\end{equation}
Here, $h_v$ is the velocity-dependent heavy quark field in HQET
\cite{Geor}, and the states are the eigenstates of the corresponding
effective Lagrangian. Details of the derivation of (\ref{Sdef}), as
well as the extension to the case of $\bar B\to X_c\,\ell\,\bar\nu$
decays, will be given elsewhere \cite{future}.

Expanding our result in powers of $1/m_b$, we obtain
\begin{eqnarray}\label{Syexp}
   S(y) &=& \sum_{n=1}^\infty\,{1\over n!}\,\delta^{(n-1)}(1-y)\,
    \bigg({2\over m_b}\bigg)^n\,(v-\hat p)_{\mu_1}\ldots
    (v-\hat p)_{\mu_n}\,\langle\,i D^{\mu_1}\ldots iD^{\mu_n}
    \rangle \nonumber\\
   &&\mbox{}+~\hbox{less singular terms.}
\end{eqnarray}
The forward scattering matrix elements between $B$ mesons, or between
any other hadronic states that are unpolarized, can be parameterized
in the form
\begin{equation}\label{Andef}
   \langle\,i D^{\mu_1}\ldots iD^{\mu_n} \rangle
   = A_n\,v^{\mu_1}\ldots v^{\mu_n}
   +~\hbox{terms with $g^{\mu_i\mu_j}$.}
\end{equation}
Since $(v-\hat p)^2=(1-y)$ vanishes at the endpoint, only the
coefficients $A_n$ contribute to the most singular terms in $S(y)$.
For the same reason, we can replace factors $2 v\cdot(v-\hat p)=
(2-y)$, which arise upon contraction of the indices in (\ref{Syexp}),
by 1. This leads to the following expression for the shape function:
\begin{equation}
   S(y) = \sum_{n=1}^\infty\,{1\over n!}\,{A_n\over m_b^n}\,
   \delta^{(n-1)}(1-y) +~\hbox{less singular terms.}
\end{equation}
{}From a comparison with (\ref{singular}) and (\ref{Mnexp}), we
obtain for the moments of $S(y)$:
\begin{equation}\label{MnAnrel}
   M_n = {1\over(n+1)}\,{A_{n+1}\over m_b^{n+1}} \,,\qquad
   a_n = {A_{n+1}\over n+1} \,.
\end{equation}
The first three coefficients $A_n$ are given by $A_0=1$, $A_1=0$, and
$A_2=-\lambda_1/3$.

\section{A model calculation}
\label{sec:5}

It is instructive to consider a simple model for the shape function
$S(y)$. For this purpose, we evaluate the expectation value in
(\ref{Sdef}) following the phenomenological approach of Altarelli et
al.\ (ACM) \cite{ACM}. We emphasize, however, that this is mainly
meant as an illustrative example rather than a prediction of the
physical shape function. In fact, we will see very clearly the
limitations and shortcomings of the ACM approach.

In the ACM model, one assumes the validity of the parton model and
incorporates bound state effect by assigning a momentum distribution
$\phi(|\vec p_b|)$ to the heavy quark inside the $B$ meson at rest.
It is then appropriate to replace the covariant derivative in
(\ref{Sdef}) by the spatial components of the heavy quark momentum
$\vec p_b$. The gluon field in the covariant derivative is neglected.
Accordingly, in the rest frame of the $B$ meson, one makes the
replacement
\begin{equation}
   {2\over m_b}\,(v-\hat p)\cdot i D
   \to {2\,\vec p_\ell\cdot\vec p_b\over m_b^2}
   = {y\,|\vec p_b|\over m_b}\,\cos\vartheta \,,
\end{equation}
where $\vartheta$ is the angle between the lepton and the heavy quark
momentum. Since we are interested in the behavior in the endpoint
region, we can set $y=1$. The matrix element in (\ref{Sdef}) is now
replaced by an integral over the momentum distribution of the heavy
quark:
\begin{equation}
   S_{\rm ACM}(y) = \int\limits_0^\infty\!{\rm d}|\vec p_b|\,
    |\vec p_b|^2\,\phi(|\vec p_b|)\,
    \int\limits_{-1}^1\!{{\rm d}\cos\vartheta\over 2}\,
    \bigg\{ \Theta\bigg[ 1 - y + {|\vec p_b|\over m_b}\,
    \cos\vartheta\bigg] - \Theta(1-y) \bigg\} \,.
\end{equation}
It is straightforward to calculate the moments of this model shape
function, and from (\ref{MnAnrel}) the corresponding predictions for
the hadronic matrix elements $A_n$. We find that $A_{2n+1}^{\rm ACM}
=0$, and
\begin{equation}
   A_{2n}^{\rm ACM} = \langle\,|\vec p_b|^{2n}\rangle
   = \int\limits_0^\infty\!{\rm d}|\vec p_b|\,|\vec p_b|^{2(n+1)}\,
   \phi(|\vec p_b|) \,.
\end{equation}
In the ACM model, one assumes a gaussian distribution,
\begin{equation}
   \phi(|\vec p_b|) = {4\over\sqrt{\pi}\,p_{\rm F}^3}\,
   \exp\bigg(-{|\vec p_b|^2\over p_{\rm F}^2}\bigg) \,,
\end{equation}
where $p_{\rm F}$ is the Fermi momentum. This leads to the shape
function
\begin{equation}
   S_{\rm ACM}(y) = \bigg[\,{1\over 2} - \Theta(1-y)\bigg]\,
   \Phi\bigg({m_b\over p_{\rm F}}\,|y-1|\bigg) \,.
\end{equation}
$\Phi(x)=\frac{2}{\sqrt{\pi}}\int\limits_x^\infty\!{\rm d}t\,
e^{-t^2}$ denotes the complementary error function. For the
coefficients $A_{2n}$, one obtains
\begin{equation}
   A_{2n}^{\rm ACM} = {(2n-1)!!\over 2^n}\,p_{\rm F}^{2n} \,.
\end{equation}
Comparing this to the general relation $A_2=-\lambda_1/3$, we derive
\begin{equation}
   -\lambda_1^{\rm ACM} = {3\over 2}\,p_{\rm F}^2\simeq
   0.08~{\rm GeV^2,}
\end{equation}
where we have used $p_{\rm F}\simeq 230$ MeV as obtained from the
most recent fit of the ACM model to experimental data \cite{Cassel}.
The model thus predicts a rather small value of $-\lambda_1$.

In Fig.~\ref{fig:3}, we show the shape function of the ACM model for
three different values of the Fermi momentum. We stress again that
this simple model calculation is presented for pedagogical purposes
only. In particular, note that in the ACM model the even moments of
the shape function (corresponding to odd coefficients $A_{2n+1}$)
vanish by rotational invariance. This is a consequence of the fact
that one replaces the {\sl operator\/} of the covariant derivative by
a $c$-number momentum vector. Whereas in QCD the commutator of two
covariant derivatives gives the gluon field strength tensor, in the
ACM model this commutator vanishes. However, exactly those terms
involving the gluon field are responsible for an asymmetry in the
shape function around $y=1$. To see this, consider the matrix element
of three covariant derivatives. Using the equations of motion of
HQET, it is easy to show that
\begin{equation}
   \langle\,i D^\mu\,iD^\nu\,iD^\alpha\,\rangle
   = A_3\,(v^\mu v^\alpha - g^{\mu\alpha})\,v^\nu \,,
\end{equation}
and taking the antisymmetric combination in $\mu$ and $\nu$, we find
that $A_3$ is related to a matrix element involving the gluon field
strength tensor:
\begin{equation}
   \langle\,ig_s\,G^{\mu\nu}\,iD^{\alpha}\,\rangle
   = A_3\,(v^\mu g^{\nu\alpha} - v^\nu g^{\mu\alpha}) \,.
\end{equation}
In QCD, there is no reason why such a matrix element should vanish.
Hence, we expect an asymmetry of the physical shape function. The
above example is instructive since it shows that a measurement of the
moments of the shape function can provide quite fundamental
information about the dynamical properties of the theory of strong
interactions.

\section{Summary and conclusions}
\label{sec:6}

We have presented a QCD-based approach to the inclusive lepton energy
spectrum in $\bar B\to X_u\,\ell\,\bar\nu$ decays. We have introduced
the concept of a shape function, which is a genuinely nonperturbative
object that describes the rapid fall-off of the spectrum in the
endpoint region. The moments of this function obey a very simple
relation to forward scattering matrix elements of local,
higher-dimension operators. QCD predicts that the leading
contribution to the first moment vanishes. The second moment, which
is a measure of the size of the endpoint region, is proportional to
the expectation value of the kinetic energy of the heavy quark inside
the hadron.

Our approach goes beyond previous work on the lepton spectrum
\cite{Bigi,Blok,MaWe}, which was applicable only for lepton energies
not too close to the endpoint. It aims at a systematic use of the
rich source of information contained in the endpoint region. Whereas
the main part of the spectrum is determined by kinematics and only
receives small nonperturbative corrections, the behavior close to the
endpoint is characterized by an infinite set of hadronic matrix
elements. It is worth noting that, although there are no first-order
power corrections to the spectrum and decay rate at small lepton
energies, there exists a small region where the true spectrum is very
different from the theoretical prediction.

There are obvious improvements of the analysis presented here. Before
confronting our results with data, it is necessary to include
radiative corrections. For the case of $\bar B\to X_u\,\ell\,\bar\nu$
decays, they are known to affect the parton model spectrum in a
significant way \cite{Cabi,ACM}. Such corrections will affect the
form of the shape function, too. We expect small perturbative
corrections of order $\alpha_s(m_b)$ to all moments of the shape
function. Moreover, radiative corrections will wash out the step in
the shape function at $y=1$, resulting in a rapidly varying, but not
discontinuous, behavior. Another important generalization of our
approach is that to the case of $\bar B\to X_c\,\ell\,\bar\nu$
decays. The nonvanishing mass of the charm quark will lead to
technical complications, but conceptually there is no problem in
defining a shape function $S(y,\rho)$ for any value of
$\rho=m_c^2/m_b^2$. The (appropriately defined) moments of this
generalized shape function are still related to the same hadronic
matrix elements as in the case of a massless final state quark. Work
on all these issues is in progress and will be reported elsewhere
\cite{future}.

We believe that our approach will eventually lead to a better
understanding of the nonperturbative aspects of inclusive decay
spectra, with implications for the measurement of some of the
fundamental parameters of the standard model, such as the heavy quark
masses and the elements $V_{cb}$ and $V_{ub}$ of the
Kobayashi-Maskawa matrix. Currently, the most promising applications
of the method seem to be the following:

In $B$ decays into charmless final states, an understanding of the
endpoint region is crucial for a reliable determination of $V_{ub}$.
The approach of Refs.~\cite{Bigi,Blok,MaWe} cannot be used for this
purpose, since current measurements are limited to a small energy
range $2.3\le E_\ell\le 2.6$ GeV (corresponding to $0.96\le y\le
1.08$ for $m_b=4.8$ GeV), which is too close to the endpoint. What is
needed is some insight into the nonperturbative effects relevant to
the shape of the spectrum in the endpoint region. An expansion in
powers of $1/m_b$ is not suitable for such a situation. The relevant
physics is encoded in the moments of the shape function, which are
related to forward scattering matrix elements of local,
higher-dimension operators. Such matrix elements can be addressed
using nonperturbative techniques such as lattice gauge theory or QCD
sum rules. It may even be possible in these approaches to attempt a
direct calculation of the shape function from its definition in
(\ref{Sdef}).

For $\bar B\to X_c\,\ell\,\bar\nu$ transitions, the situation is very
different. Already, there exist very accurate measurements of the
lepton spectrum in this case. For a determination of $V_{cb}$ and of
the quark masses $m_b$ and $m_c$, an understanding of the endpoint
region is thus not a necessary requirement. However, these decays
offer the exciting possibility to extract the shape function from the
data, simply by subtracting the (corrected) parton model spectrum
from the measured distribution. One could then compute the first few
moments of the shape function and extract some of the coefficients
$A_n$, which encode fundamental dynamical properties of QCD. The fact
that QCD predicts the size of the first moment $M_0$ in
(\ref{M0M1val}) provides an important constraint, which can help to
obtain a very precise determination of the $b$-quark mass.

We conclude with a speculation about yet another exciting
possibility, namely to combine the analyses of $\bar B\to
X_c\,\ell\,\bar\nu$ and $\bar B\to X_u\,\ell\,\bar\nu$ decays. In
fact, one can derive a simple relation between the shape functions
for these two processes \cite{future}. One can imagine to measure
the shape function in $B$ decays into charmed particles, and then
predict the shape function for charmless transitions. This avenue may
well be the most promising one with respect to a precise extraction
of $V_{ub}$.

\bigskip\bigskip
{\it Acknowledgments:\/}

It is a pleasure to thank Thomas Mannel for useful discussions and
for collaboration on subjects related to this work.

\newpage

\begin{center}
\Large\bf Figure Captions
\end{center}

\begin{figure}[h]
\caption{\label{fig:1}
An example of a resummed lepton energy spectrum according to
(\protect{\ref{resum}}). On the vertical axis, we show ${\rm d}
\Gamma/{\rm d}y$ in units of $2\Gamma_b$.}
\end{figure}

\begin{figure}[h]
\caption{\label{fig:2}
Two shape functions with identical moments $M_0$ and $M_1$ (a), and
the corresponding lepton spectra (b).}
\end{figure}

\begin{figure}[h]
\caption{\label{fig:3}
The shape function of the ACM model (a) and the corresponding lepton
spectrum (b) for $-\lambda_1=0.05~{\rm GeV^2}$ (dashed), 0.1 GeV$^2$
(solid), and 0.2 GeV$^2$ (dotted). The corresponding values of the
Fermi momentum are $p_{\rm F}\simeq 180$ MeV, 260 MeV, and 365 MeV,
respectively. In (b), the parton model spectrum is shown as a grey
line.}
\end{figure}

\end{document}